# Inconsistency identification in network meta-analysis via stochastic search variable selection

Georgios Seitidis*[1], Stavros Nikolakopoulos[2,3], Ioannis Ntzoufras[4], Dimitris Mavridis[1]

[1]Department of Primary Education, University of Ioannina, Ioannina, Greece

[2]Department of Psychology, University of Ioannina, Ioannina, Greece

[3]Department of Biostatistics, Julius Center for Health Sciences and Primary Care, University Medical Center Utrecht, Utrecht, The Netherlands

[4]Department of Statistics, Athens University of Economics and Business, Athens, Greece

*Corresponding author; g.seitidis@uoi.gr

**Abstract**

The reliability of the results of network meta-analysis (NMA) lies in the plausibility of key assumption of transitivity. This assumption implies that the effect modifiers' distribution is similar across treatment comparisons. Transitivity is statistically manifested through the consistency assumption which suggests that direct and indirect evidence are in agreement. Several methods have been suggested to evaluate consistency. A popular approach suggests adding inconsistency factors to the NMA model. We follow a different direction by describing each inconsistency factor with a candidate covariate whose choice relies on variable selection techniques. Our proposed method, Stochastic Search Inconsistency Factor Selection (SSIFS), evaluates the consistency assumption both locally and globally, by applying the stochastic search variable selection method to determine whether the inconsistency factors should be included in the model. The posterior inclusion probability of each inconsistency factor quantifies how likely is a specific comparison to be inconsistent. We use posterior model odds or the median probability model to decide on the importance of inconsistency factors. Differences between direct and indirect evidence can be incorporated into the inconsistency detection process. A key point of our proposed approach is the construction of a reasonable “informative” prior concerning network consistency. The prior is based on the elicitation of information derived historical data from 201 published network meta-analyses. The performance of our proposed method is evaluated in two published network meta-analyses. The proposed methodology is publicly available in an R package called *ssifs*, developed and maintained by the authors of this work.

## 1. Introduction

In evidence synthesis, Network Meta Analysis (NMA) is typically employed to evaluate the relative efficacy and/or safety of multiple available interventions[1–3]. A key assumption in NMA is that of transitivity, also termed in the literature as the similarity assumption, which refers to the ability to learn about a treatment effect indirectly[3]. NMA synthesizes direct and indirect evidence and violations of transitivity may manifest themselves in large differences between these two sources of evidence, also known as inconsistency in the network.

Several methods have been suggested to evaluate the consistency assumption that can be broadly categorized into local and global methods. Local methods evaluate consistency either in every loop or comparison in the network for which we have both direct and indirect evidence[4–9]. Global methods include approaches where inconsistency factors are added in the NMA model and evaluated simultaneously[10–12], employing the generalized Cochran's Q statistic[13], fitting the NMA model using integrated nested Laplace approximations[14], and treating the comparisons as independent parameters for estimation[15]. Local methods have an increased probability of false positives for large networks, while global methods suffer from lack of power[10]. Moreover, local methods in which for each comparison we contrast direct evidence and indirect effect estimates (node-splitting, symmetric side-splitting, separate indirect from direct design evidence SIDDE), have been shown in the presence of multi-arm trials to disregard the principle of independence between direct and indirect evidence and as a result, may lead to false conclusions[8].

The Lu and Ades model is a popular method for testing the consistency assumption globally, which is evaluated by adding inconsistency factors to each independent closed-loop[12]. Drawbacks of the method are: 1) the model is not uniquely defined and different parameterisations (adding an inconsistency factor to different comparisons within a loop) may lead to different results, 2) it accounts only for loop inconsistencies, and 3) in the presence of multi-arms studies the implementation of the method is not straight-forward. Until now, the implementation of the method has been carried out manually, by specifying the inconsistency factors by hand. Although this may not be an issue in small networks with few loops, in complex networks, the specification of the inconsistency factors is extremely hard, making the implementation of the method in most cases not feasible.

When employing NMA in a Bayesian framework, inconsistency models have been used to relax the consistency assumption by treating inconsistency as an additional random effect[10–12]. Popular methods for exploring inconsistency include estimating the variance of the inconsistency random effects, computing deviance contribution and comparing the deviance information criterion (DIC) for consistency and inconsistency models[15]. It should be noted that both consistency and inconsistency models are essentially linear regression models. In the inconsistency model, extra inconsistency factors are added as covariates to account for any difference between direct and indirect evidence. Therefore, by employing variable selection techniques, the importance of inconsistency factors and the plausibility of the consistency assumption can be thoroughly assessed.

The scope of this article is two-fold: Firstly, we propose employing a Bayesian variable selection approach tailored to the NMA inconsistency model, aiming to evaluate the consistency assumption both locally and globally. Subsequently, we present a consistent algorithm for specifying the inconsistency factors in the Lu and Ades model. For the implementation of the suggested methodology, we developed the R package *ssifs*[16]. The article is structured as follows. We first describe the NMA random-effects model and how it is formulated to include inconsistency factors. Next, we present our proposed method, and we illustrate it in two datasets from the field of medicine. The article concludes with a discussion about possible extensions of the method.

## 2. Methods

### 2.1. The NMA consistency and inconsistency models

Suppose that we have consider trials that evaluate the efficacy of three treatments A, B and C (ABC network). Further, assume that all studies under consideration provide information for the comparison between all three pairs of interventions forming a closed loop of evidence. The indirect estimate (denoted with an upper index "*Ind*") for a comparison (say BC) would be $\mu_{BC}^{Ind} = \mu_{AC}^{Dir} - \mu_{AB}^{Dir}$ where upper index "*Dir*" denotes direct evidence. Consistency implies that parameters measuring direct and indirect evidence should be equal for all treatment comparisons. Hence, for the comparison between treatments B and C we should have

$$\mu_{BC}^{Dir} = \mu_{BC}^{Ind} \Leftrightarrow \mu_{BC}^{Dir} = \mu_{AC}^{Dir} - \mu_{AB}^{Dir}. \quad \mathbf{(1)}$$

Equation 1 is fundamental for the NMA consistency model. If we have T interventions, it suffices to determine $\mathrm{T} - 1$ basic contrasts that include all competing interventions. Usually, we assume a reference intervention (say A) and the basic contrasts are AB, AC, etc. Any treatment effect can either belong in the basic contrasts or written as a function of those using Equation 1.

Suppose that we have a collection of $n$ studies where each study $s$ compares $T_s$ interventions, $s = 1, 2, \dots, n$. If $T_s = 2$ the study is called "two-arm", otherwise it is called "multi-arm". The NMA model requires $T_s - 1$ comparisons from each study since the rest can be obtained as a linear combination, yielding a total of $N = \sum_{s=1}^{n}(T_s - 1)$ comparisons to be included in the NMA model. Let the vector $\boldsymbol{y} =$ $(y_1, \ y_2, \ldots, \ \ y_N\,)'$ contain the estimated contrasts across all studies. Hence, $y_i$ denotes the $u_{is} \in \{1, \ldots, T_s - 1\}$ contrast of study $s \in \{1, 2, \ldots, n\}$. The random-effects NMA model can be written as

$$\boldsymbol{y} = \boldsymbol{X\mu} + \boldsymbol{\beta} + \boldsymbol{\epsilon}$$

where $\boldsymbol{X}$ is the $N \times (T - 1)$ design matrix which uses the basic contrasts as columns and includes the comparisons observed in each study embedding the consistency equations, $\boldsymbol{\mu}$ is a vector of length $T - 1$ with elements the basic contrasts, $\boldsymbol{\beta}$ is a vector of normally distributed random-effects $\boldsymbol{\beta} \sim MN_N(\boldsymbol{0}, \ \boldsymbol{\Delta})$ of dimension $N$ and $\boldsymbol{\epsilon}$ is a vector of normally distributed sampling errors $\boldsymbol{\epsilon} \sim MN_N(\boldsymbol{0}, \ \boldsymbol{\Sigma})$ of dimension $N$. Each study effect $\mu_j$ is recorded as a column in matrix $\boldsymbol{X}$ denoted by $\boldsymbol{X}_j$. For each observation $y_i$ concerning study $s$ and comparison $u_{is}$ between treatments $t_{u_{is}}^{(1)}$ and $t_{u_{is}}^{(2)}$, we set that $X_{ij} = 1$ if $j$ comparison is the same as $u_{is}$, or $X_{ij} = -1$ if the $j$ comparison is in the opposite direction. When comparison $u_{is}$ is not among the basic contrasts, then the linear predictor should accommodate Equation 1. Hence, for a comparison $u_{is}$, when neither treatment $t_{u_{is}}^{(1)}$ nor $t_{u_{is}}^{(2)}$ is the reference treatment, then we set $X_{ij} = 1$ if $j$ comparison refers to the contrast of the reference treatment versus treatment $t_{u_{is}}^{(2)}$, $X_{ij} = -1$ if $j$ comparison refers to the contrast of the reference treatment versus treatment $t_{u_{is}}^{(1)}$, and $X_{ij} = 0$ for the rest basic contrasts. For example, for an ABCD network with A as a reference treatment, if $u_{is} =$ BC, then we set $X_{i1} = -1, \ X_{i2} = 1$ and $X_{i3} = 0$ corresponding to basic contrasts AB, AC, and AD, respectively. Finally, covariance matrix $\boldsymbol{\Delta}$ is a block diagonal matrix, assuming common heterogeneity across treatment comparisons, while covariance matrix $\boldsymbol{\Sigma}$ is assumed known and is estimated from the data.

Inconsistency arises when a significant disagreement between direct and indirect evidence is observed, for example in our ABC network when $\mu_{BC}^{Dir}$ differs substantially from $\mu_{BC}^{Ind}$. In such a case, inconsistency factor $b_{ABC}$ could be added in the consistency equation to relax this assumption, which is now expressed mathematically as

$$\mu_{BC}^{Ind} = \mu_{AC}^{Dir} - \mu_{AB}^{Dir} + b_{ABC}.$$

The consistency assumption could be tested by evaluating whether the inconsistency factor $b_{ABC}$ differs significantly from zero. In the general case, where $p$ inconsistency factors are added in the NMA model, the model has the following form

$$\boldsymbol{y} = \boldsymbol{X\mu} + \boldsymbol{\beta} + \boldsymbol{Zb} + \boldsymbol{\epsilon} \qquad (\mathbf{2})$$

where $\boldsymbol{Z}$ is an $N \times p$ index matrix with elements values of 1, 0, -1 indicating in which comparisons inconsistency factors are added in the linear predictor, and $\boldsymbol{b} = (b_1, b_2, \dots, b_p)'$ is a vector with the coefficient of each inconsistency factor.

### 2.2. Specification of Inconsistency factors

In NMA, we have two types of inconsistency: (a) loop and (b) design inconsistency[10].

In order to specify the first, we need first to introduce the notion of loop and closed-loop in a network. The term “loop” is used to specify a path formed by a group of interventions. A loop is closed when a group of interventions is connected in the network graph with a polygon. In practice it means that we can calculate all pairwise treatment comparisons either directly or indirectly. Loop inconsistency is present, when in a closed-loop, direct and indirect evidence vary substantially. For example, in a three-treatment loop, if Equation 1 does not hold, we have loop inconsistency. Multi-arm trials are by definition consistent complicating the definition of loop inconsistency.

The second type, design inconsistency, was introduced by Higgins et al[10] . The term design refers to the set of treatments compared within a study. Design inconsistency refers to differences in effect sizes between studies involving different sets of interventions. For example, suppose that in the previous ABC network, ABC three-arm studies are also present, and the effect estimate of comparison AB between the AB and ABC studies differs substantially.

#### 2.2.1. Specification of Inconsistency Structure

Several models have been developed for the specification of ***Z*** matrix which is the design matrix specifying which inconsistency parameters should be added in the linear predictor (2). Lu and Ades proposed a model which accounts only for loop inconsistencies in the networks, and the consistency assumption is tested in every independent closed-loop[12]. In complex networks, the implementation of the method is challenging since the specification of the inconsistency factors is sequentially performed manually. Thus, we have developed an algorithm that automatically specifies the comparisons where inconsistency factors should be added. Figure 1 presents the process flowchart of the algorithm. More details about the algorithm can be found in the appendix.

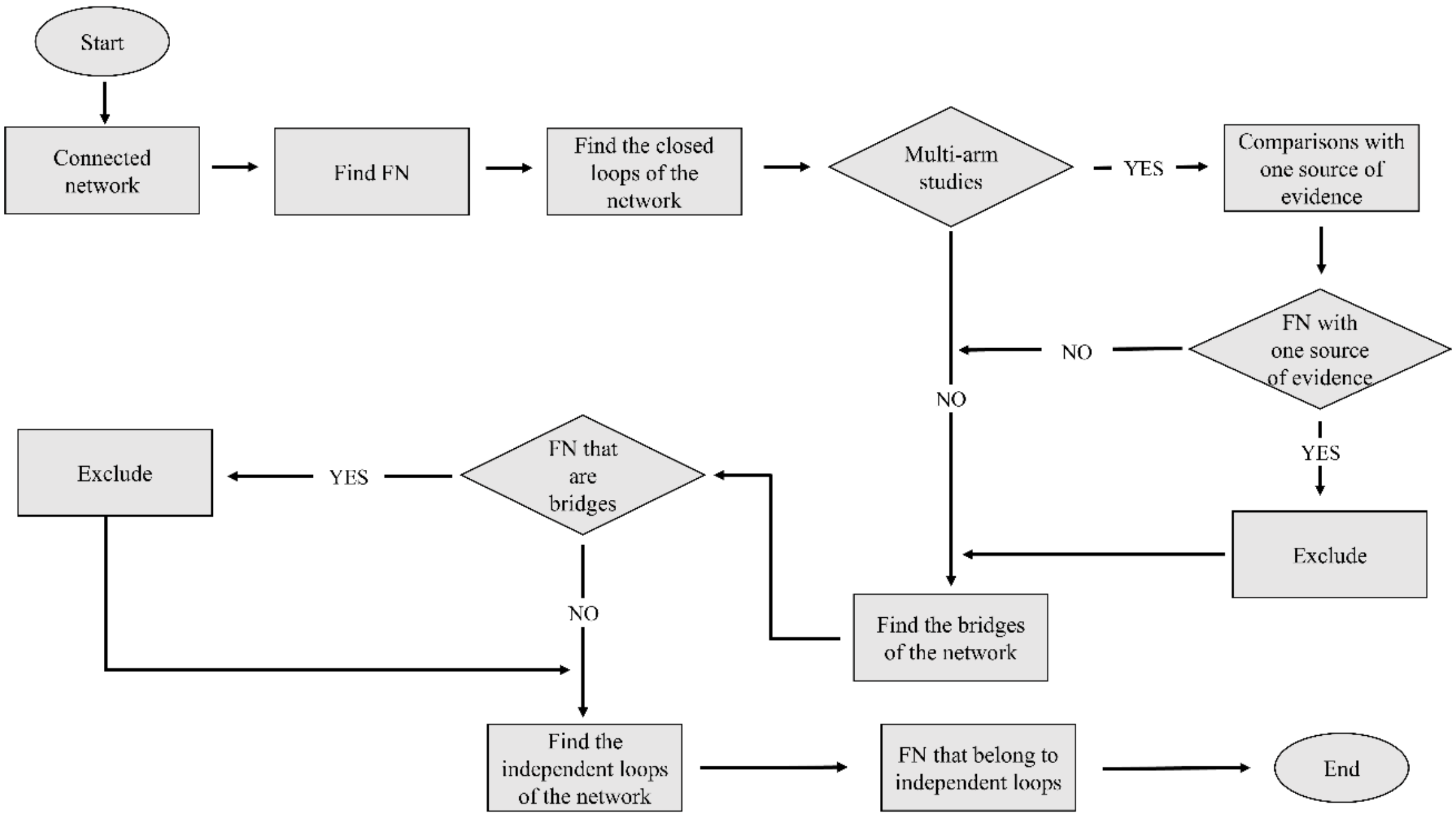

**Figure 1:** Process flowchart of identifying the comparisons in which inconsistency factor should be added based on the Lu and Ades model. (Abbreviation FN: functional parameters)

The design-by-treatment interaction model[10] was proposed by Higgins et al as a method that accounts for both types of inconsistencies. Moreover, Jackson et al developed a random-effects implementation of the design-by-treatment interaction model, in which consistency can be evaluated at each design of the network[11].

### 2.3. Stochastic Search Inconsistency Factor Selection

Concerning the evaluation of the network inconsistencies, we propose an implementation of the Stochastic Search Variable Selection (SSVS) algorithm[17] which will be referred in the following as Stochastic Search Inconsistency Factors Selection (SSIFS). SSIFS is a two-step method in which, at the first step, we add inconsistency factors in the NMA model based on the specification method of matrix $\boldsymbol{Z}$ described in Section 2.2.1. At the second stage, we treat the detection of inconsistency factors as a variable selection problem and use SSVS to identify inconsistencies in a network under study.

After specifying $\boldsymbol{Z}$, following the SSVS model specification, each element of $\boldsymbol{b}$ is modelled as a realization from a mixture of two normal distributions with different variances. A latent vector $\boldsymbol{\gamma} = (\gamma_1, \gamma_2, \dots, \gamma_p)$ is associated with each element of $\boldsymbol{b}$ which operates as an indicator variable (taking values 0 or 1) identifying whether the respective element of $\boldsymbol{b}$ lies in a small area around zero or not. We represent the normal mixture by

$$b_\ell \,|\gamma_\ell \sim (1-\gamma_\ell)N(0, \psi_\ell^2) + \gamma_\ell N(0, c^2\psi_\ell^2)$$

where $\ell = 1, 2, \dots, p$. The first density ("spike") denotes the effect of the inconsistency factor when it is minor (i.e. close to zero) and therefore it should not be included in the model ($\gamma_\ell = 0$), while the second ("slab"), denotes the inconsistency factor effect when it away from zero and therefore it should be included in the model ($\gamma_\ell = 1$). The spike and the slab prior setup is depicted in Figure 2.

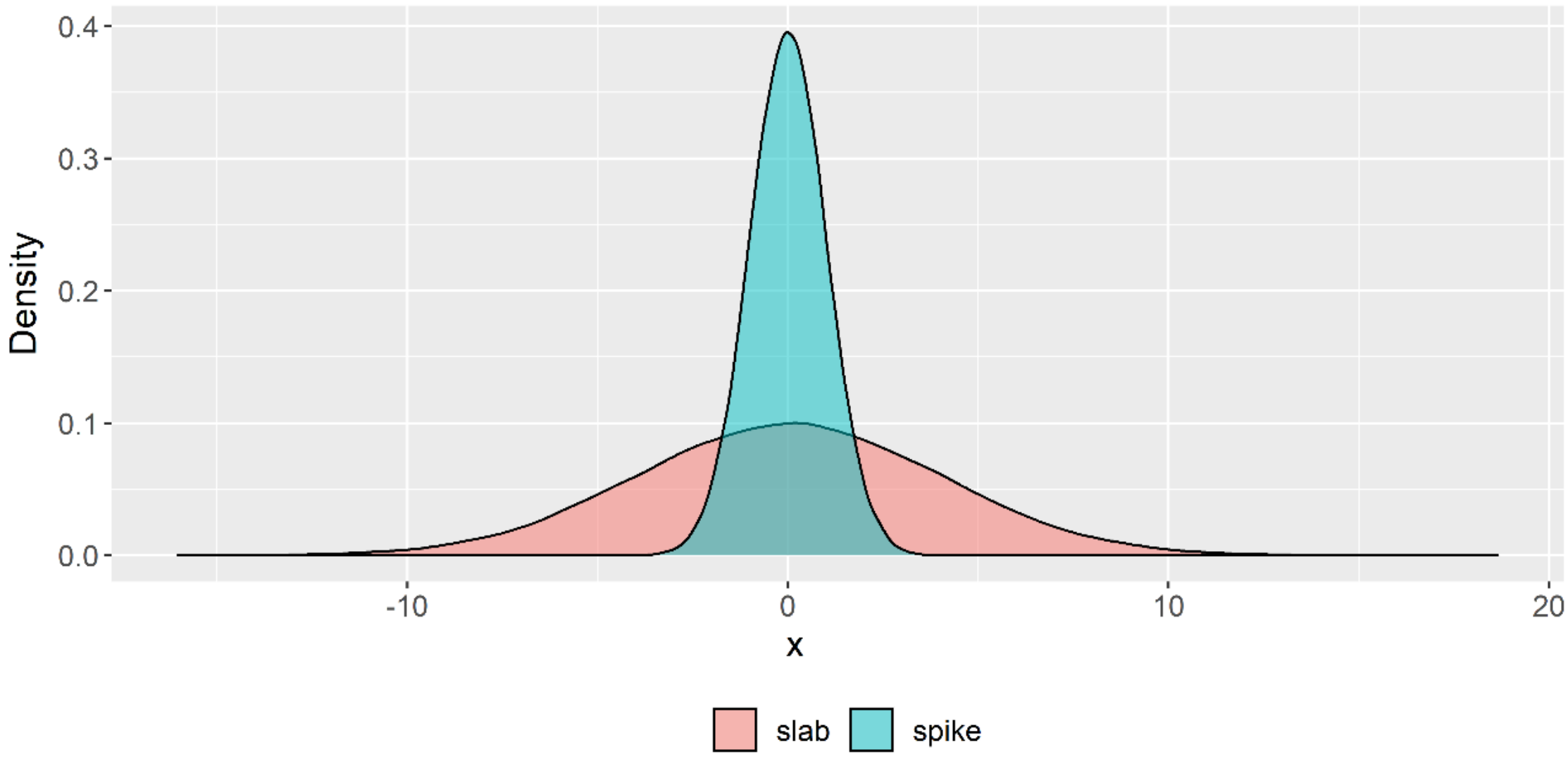

**Figure 2:** Graphical representation of the SSVS Spike and slab prior setup.

The mixture of these two densities can be represented in a condensed matrix notation by

$$\boldsymbol{b}|\boldsymbol{\gamma} \sim \boldsymbol{N}\left(\boldsymbol{0}, \boldsymbol{D}_{\boldsymbol{\gamma}}\boldsymbol{R}\boldsymbol{D}_{\boldsymbol{\gamma}}\right)$$

where $\boldsymbol{R}$ is the prior correlation matrix of the elements in $\boldsymbol{b}$, $\boldsymbol{D}_{\boldsymbol{\gamma}} = diag(a_1\psi_1, \dots, a_p\psi_p)$ with $a_\ell = c^\gamma$ (i.e. $a_\ell = 1$ if $\gamma_\ell = 0$ and $a_\ell = c$ if $\gamma_i = 1$); where $\psi_\ell$ and $c$ are tunning parameters controlling the mixing ability of the method, the penalty induced for each extra parameter included in the model and the interval of non-importance for each inconsistency.

#### 2.3.1. Prior setup

A convenient choice for matrix $\boldsymbol{R}$ is to assume that is equal to the identity matrix $\boldsymbol{I}_p$. This setting implies that the inconsistency factors are independent, and the inclusion of an inconsistency factor does not affect the inclusion of another. In a network of interventions where each effect estimate affects all other estimates, independence of inconsistency factors is questionable. A popular choice for prior setup in Bayesian variable selection is the Zellner's g-prior[18]. Under this approach, the correlation matrix $\boldsymbol{R}$ is written by the following expression

$$\boldsymbol{R} = g(\boldsymbol{Z}^{\boldsymbol{T}}\boldsymbol{Z})^{-1}\sigma^2,$$

while we consider an improper prior for the error variance of the type

$$\pi(\sigma^2) \propto \frac{1}{\sigma^2}.$$

The main advantages of the Zellner's g-prior are that the general correlation of $\boldsymbol{b}$ is encapsulated by the matrix $(\boldsymbol{Z}^{\boldsymbol{T}}\boldsymbol{Z})^{-1}$ while we have only one remaining variance parameter to specify in comparison with the general formulation where we need to specify the whole matrix $\boldsymbol{R}$. Nevertheless, the specification of $g$ should be cautious in order not to trigger the Jeffreys-Lindley-Bartlett paradox[19–21], in which for proper large values of $g$ the posterior odds tend to favour the simplest model, which in our case is the consistency model. For this reason, several approaches have been proposed, including the use of hyper-prior[22,23] or assuming $g$ to be fixed. We use the common choice of unit-information criteria approach[24] which mathematically translates to $g = N$. The error variance term $\sigma^2$ assumed to follow the Jeffreys scale-invariant prior. An alternative for the error variance is to assume a non-informative inverse-gamma prior with equal shape and scale parameters, typically set to $10^{-3}$ or $10^{-4}$ [25]. This choice has the advantage of being proper. Nevertheless, the results are not sensitive to the choice of this prior or parameters provided that these values are large enough.

### 2.3.2. Prior inclusion probabilities

Prior inclusion probabilities define the probability of including each inconsistency factor in the model, a priori. It is crucial to specify these probabilities appropriately in order to avoid issues with multiple comparisons. The standard but naïve non-informative choice of equal prior probabilities for all models does not account for multiplicity[26]. A standard approach in Bayesian variable selection is to use a Beta-binomial distribution on the prior inclusion probabilities, which accounts for multiplicity by automatically introducing a penalty that handles multiple testing[26,27]. This expressed mathematically as

$$\begin{gathered} \gamma_\ell \sim Bernoulli(\pi) \\ \pi \sim Beta(\alpha, \beta). \end{gathered} \qquad \textbf{(3)}$$

In the simple case where $\alpha = \beta = 1$, this is equivalent to assigning a uniform prior to π. Even though this setting appears appealing, it cannot be applied to SSIFS as it will unrealistically favor inconsistent NMA models. Consider the simple case where we assume that each inconsistency factor is Bernoulli distributed with 0.5 inclusion probability

$$\gamma_\ell \sim Bernoulli(0.5), \qquad \ell = 1, 2, \dots, p.$$

Let $\pi_\ell = P(\gamma_\ell = 1)$ denote the probability that the inconsistency factor is included in the model and $1 - \pi_\ell = P(\gamma_\ell = 0)$ the probability not included. The prior probability $\pi_{cons}$ of the consistency model will then given by

$$\pi_{cons} = \prod_{\ell=1}^{p} P(\gamma_\ell = 0) = 0.5^p.$$

Therefore, the probability of the consistency model decreases exponentially as the number of inconsistency parameters increases. Also, this choice will tend to favor models with $\frac{p}{2}$ inconsistency factors. Alternatively, inclusion probabilities should be specified in an such manner where the probability of the consistency model remains fixed. Thus, the common inclusion probability $\pi$ for all inconsistency factors can be obtained by:

$$\pi_{cons} = P(\gamma_1 = 0, \dots, \gamma_p = 0) \Leftrightarrow \pi_{cons} = (1 - \pi)^p \Leftrightarrow$$

$$\pi = 1 - \pi_{cons}^{1/p}.$$

SSIFS can also incorporate past knowledge concerning network consistency. Parameter $\pi_{cons}$ denotes the analyst's prior belief of having a consistent network. If experts' opinion is available, an informative prior could be used. Otherwise, we consider the value of 0.5 to express our prior ignorance. In a review of 201 NMA networks, 44 networks were found to be globally inconsistent[28]. Thus, the probability to have a consistent network is estimated from this study to be equal to 0.78. This could be used as an informative prior for $\pi_{cons}$ by assuming this probability either fixed ($\pi_{cons} = 0.78$) or to follow a Beta prior distribution $\pi_{cons} \sim Beta(\alpha = 157,\ \beta = 44)$ with mean equal to 0.78 and a standard deviation of 0.03.

It should be noted that in SSIFS we model the prior inclusion probabilities of the inconsistency factors through the parameter $\pi_{cons}$. This approach is quite different from the classical approach used in Bayesian variable selection (Equation 3), where the prior on the model space is specified through the prior inclusion probabilities $\gamma_\ell$ are defined based on the parameter $\pi_\ell$.

The estimate of the posterior inclusion probability for an inconsistency factor $\ell$ is obtained as the average number of times the inconsistency factor was included in the NMA model ($\gamma_\ell = 1$) in $M$ iterations of the MCMC algorithm, that is

$$\hat{f}(\gamma_\ell = 1|\boldsymbol{y}) = \frac{1}{M-B}\sum_{t=B+1}^{M} I(\gamma_\ell^{(t)} = 1).$$

where $B$ is the number of iterations considered as burn-in period, and $\gamma_\ell^{(t)}$ is the inclusion value of $\ell$ inconsistency factor at iteration $t$.

### 2.3.3. Tunning

Tunning of SSIFS parameters is essential in order to ensure a good mixing of the MCMC method but also in order to attained sensible results in terms of model comparison. The setting of $\psi_\ell$ and $c$ must be in such a way that $\psi_\ell^2$ be small and $c^2\psi_\ell^2$ large. Possible values of $\psi_\ell$ could be the inconsistency factor's standard deviations obtained from a pilot MCMC run of the full NMA model[29]. Parameter $c$ can be interpreted as the prior odds of excluding the inconsistency factor when its effect is close to zero. The larger the value of $c$, the stricter we are about including inconsistency factors with small effects. It has been observed that values between 10 and 100 usually perform well in most cases however caution may be needed depending on the data characteristics[17,30].

In order to specify the prior parameters of the variable selection part ($\psi_\ell$ and $c$), we can consider a priori assume a minimum accepted value (say $\omega$) for the inconsistency parameters $\boldsymbol{b}$. This is referred as the level of practical significance of an inconsistency parameter. Thus, an inconsistency factor with a coefficient larger than $\omega$ in absolute values ($|b_\ell| > \omega$), should be a-priori included in the NMA model ($\gamma_\ell = 1$).

This can be achieved by setting the parameters $\psi_\ell$ and $c$ according to the intersection of the two priors used under the consistency and inconsistency status of the model; that is $f(b_\ell\,|\gamma_\ell = 0)$ and $f(b_\ell|\gamma_\ell = 1)$, respectively. The intersection point can be found by solving the equation $f(b_\ell|\gamma_\ell = 0) = f(b_\ell|\gamma_\ell = 1)$ with regards to $\boldsymbol{b}$; see Appendix for details. In the case where $\boldsymbol{R} = \boldsymbol{I}_p$, the density of $f(b_\ell|\gamma_\ell = 1)$ will be higher than the density of $f(b_\ell|\gamma_\ell = 0)$ when $|b_\ell| > \psi_\ell\sqrt{\xi(c)}$, where $\xi(c) = \frac{2c^2\, log\,(c)}{c^2-1}$ denotes how many standard deviations an inconsistency factor should be away from zero to be considered significant. For example, if a difference above 0.2 is considered important, one possible parameterization is to set $c = 10$. Then we set $0.2 = \psi_\ell\sqrt{\xi(c)}$, obtaining $\psi_\ell = \frac{0.2}{\sqrt{\xi(10)}} \approx 0.1$. In the case where a correlation between the $p$ elements of the inconsistency parameter vector $\boldsymbol{b}$ is assumed, tunning should be based on solving the following inequality

$$\boldsymbol{b}'\left[\boldsymbol{D}_{\boldsymbol{\gamma}=\mathbf{0}}(\boldsymbol{Z}'\boldsymbol{Z})^{-1}\boldsymbol{D}_{\boldsymbol{\gamma}=\mathbf{0}}\right]^{-\mathbf{1}}\boldsymbol{b} \leq gp\sigma^2\xi(c), \qquad \xi(c) = \frac{2c^2 log\,(c)}{c^2-1}$$

More details about tunning can be found in Ntzoufras et al[31] and in Mavridis et al[32].

### 2.3.4. Inconsistency Detection

In an NMA network with p inconsistency factors there are $2^p$ possible models. The aim of SSIFS is to find the most probable model in terms of posterior probability, termed as maximum a posteriori model. A common approach for model comparison in the Bayesian framework is to calculate the posterior model odds. When comparing two models $m_1$ and $m_2$, the posterior odds of $m_1$ over $m_2$ calculated as

$$PO_{m_1m_2} = \frac{f(m_1|\boldsymbol{y})}{f(m_2|\boldsymbol{y})} = \frac{f(\boldsymbol{y}|m_1)}{f(\boldsymbol{y}|m_2)}\frac{f(m_1)}{f(m_2)} = BF_{m_1m_2}\frac{f(m_1)}{f(m_2)}$$

where $BF_{m_1m_2}$ is the Bayes factor of $m_1$ over $m_2$ and $f(m_1)$ and $f(m_2)$ are the prior model probabilities. If the prior model probabilities are equal, then the posterior model odds and the Bayes factors coincide. Posterior model odds can be obtained as the ratio of the posterior model probabilities which in SSIFS is estimated by

$$\hat{f}(m|\boldsymbol{y}) = \frac{1}{M-B}\sum_{t=B+1}^{M} I(m^{(t)} = m)$$

where $m^{(t)}$ is the model indicator in $t$ iteration which transform the $\boldsymbol{\gamma}$ to a unique decimal number and calculated as

$$m(\boldsymbol{\gamma}) = \sum_{\ell=1}^{p} \gamma_\ell 2^{\ell-1}.$$

Posterior odds of the consistent NMA model $(m(\boldsymbol{\gamma}) = 0)$ over all the other observed inconsistent NMA models $(m(\boldsymbol{\gamma}) \neq 0)$ indicates if the NMA model is globally consistent.

An alternative strategy is to find the *median probability model* which includes only inconsistency factors with posterior inclusion probability larger than 0.5. For complex networks and/or too many inconsistency factors (e.g more than 20), the SSIFS may not accurately calculate the posterior model probabilities[29]. In such cases, a good strategy is to reduce the dimension by excluding the inconsistency factors that have posterior inclusion probability below 0.20 [33].

## 3. Examples

To illustrate our method, we used two examples, one on smoking cessation reported by the AHCPR Smoking Cessation Guideline Panel[12], and the other on comparative effectiveness of oral phosphodiesterase type-5 inhibitors for erectile dysfunction[28,34]. Matrix $\boldsymbol{Z}$ in SSIFS is specified using the Lu and Ades model, the design-by-treatment and the Jackson's approach. For the specification of the prior correlation matrix $\boldsymbol{R}$, we took into consideration both approaches. Firstly, we assume that inconsistency factors are independent by setting $\boldsymbol{R} = \boldsymbol{I}_p$. Secondly, that there is a dependency between inconsistency factors which is descripted by the Zellner's $g$-prior ($\boldsymbol{R} = g(\boldsymbol{Z}^{\boldsymbol{T}}\boldsymbol{Z})^{-1}\sigma^2$). Tunning was performed by assuming 0.2 as the minimum value of inconsistency that is of practical significance on the log scale. Regarding the probability of observing a consistent network, we used two different scenarios, where at the first $\pi_{cons} = 0.5$ and at the second $\pi_{cons} \sim Beta(157,\ 44)$. The first is the uniform prior on model space approach while the second is informative based on previous historical information.  A burn-in period of 50K iterations, and a total of 300K iterations were employed, using two chains to monitor convergence.

SSIFS was performed using the R package *ssifs* developed by the authors of this work. Details on the installation and the usage of package can be found on "https://georgiosseitidis.github.io/ssifs/".

### 3.1. Smoking cessation dataset

The dataset consists of 24 studies (22 two-arm studies and two three-arm studies) comparing the relative effects of no contact (reference), self-help, individual counseling and group counseling on smoking cessation. Network's geometry is presented in the Figure 3.

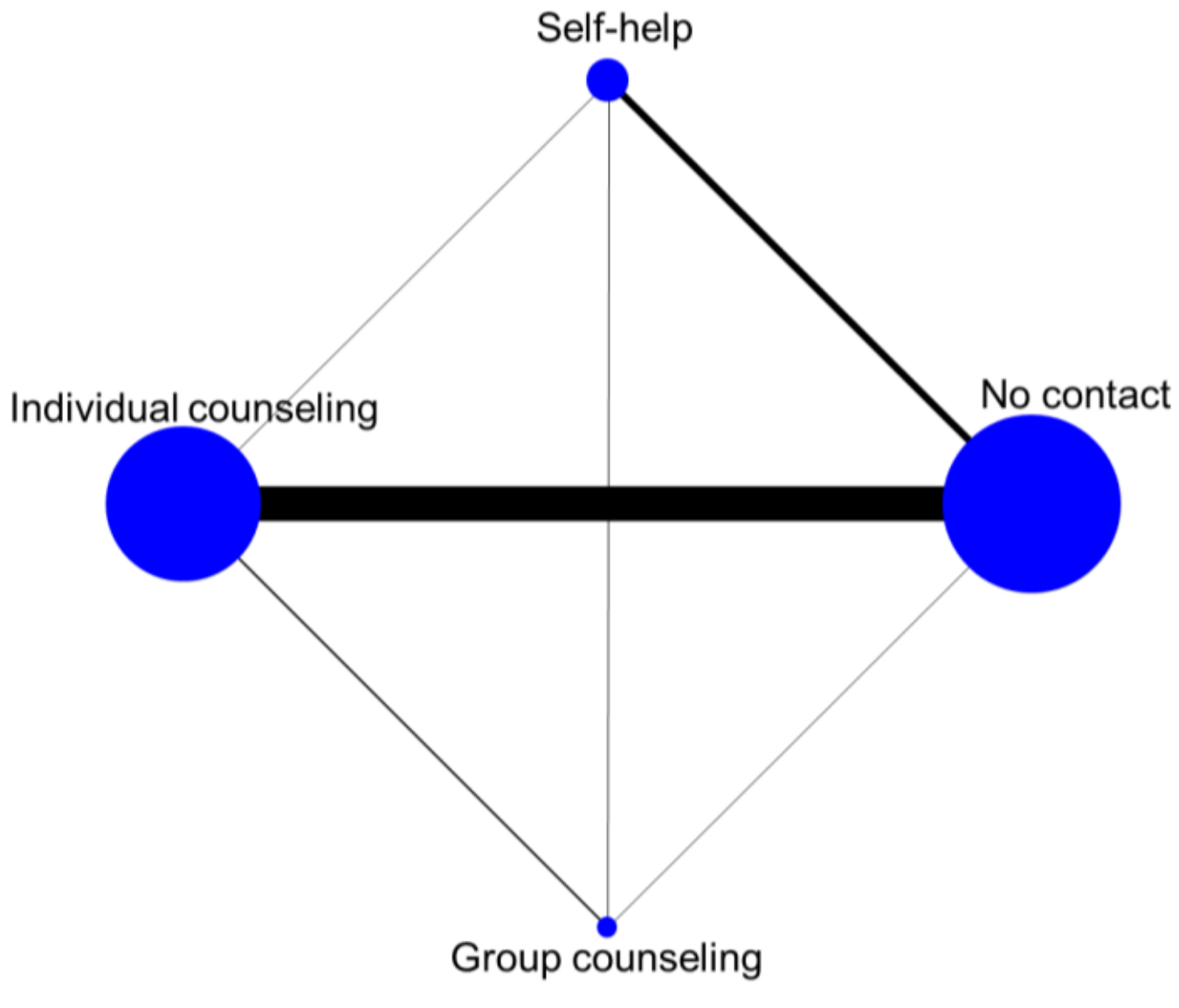

*Edge's thickness and node's size is analogous to the number of treatment comparisons and participants, respectively.*

**Figure 3:** Network geometry of smoking cessation data.

SSIFS indicates that the NMA model is consistent since, in all scenarios inconsistency factor's effects are estimated as non-important a-posteriori, yielding posterior inclusion probabilities close to zero (for more details see Figure 4 and Table S1). Hence, the median probability model clearly suggests that no important design or loop inconsistencies are present in the network. Moreover, from Table 1 we can see that the consistent NMA model (model without inconsistency factors) is estimated as maximum a-posteriori one with posterior probability around 0.56 when $\pi_{cons} = 0.5$ and above 0.81 when $\pi_{cons} \sim Beta(157, 44)$. As a result, the posterior odds favor the consistent NMA model when compared with the inconsistent NMA models (for more details see Tables S2-S13 in the supplementary material). Even if all the inconsistent NMA models are treated as a single model, the posterior odds of consistency versus inconsistency still favors the consistency hypothesis, with posterior odds ranging from 1.2 to 1.4 (marginal evidence) when the uniform prior on model space is used, and from 4.2 to 4.9 when historical information is used; see Table 1 for more details. Thus, we may safely conclude that the network is globally consistent without any significant local inconsistencies. Our findings are in agreement with previous work which also suggests that the network is consistent. In addition, Dias et al. revealed no substantial differences between the random-effects NMA model and the node-split models, indicating the absence of inconsistency[5]. Also, Lu & Ades revealed that the global model fit statistics nor the inconsistency p-value of 0.27 suggests the presence of inconsistency[12]. Nevertheless, they point out that due to the high level of heterogeneity, we cannot confidently conclude that there is no inconsistency. This is also reflected in SSIFS in the case where we used a non-informative prior for the specification of parameter $\pi_{cons}$, in which marginal evidence is available that the network is consistent.

**Table 1:** Posterior model odds ($PO$) of consistent NMA model vs inconsistent NMA models and the corresponding probability of the consistent NMA model ($\widehat{m}_{cons}$) for the three different inconsistency modelling approaches.

| Inconsistency modelling approach | $\pi_{cons} = 0.5$ | | | | $\pi_{cons} \sim Beta(157,44)$ | | | |
|---|---|---|---|---|---|---|---|---|
| | $\boldsymbol{R} = \boldsymbol{I}_p$ | | $\boldsymbol{R} = g(\boldsymbol{Z}'\boldsymbol{Z})^{-1}\sigma^2$ | | $\boldsymbol{R} = \boldsymbol{I}_p$ | | $\boldsymbol{R} = g(\boldsymbol{Z}'\boldsymbol{Z})^{-1}\sigma^2$ | |
| | $\widehat{m}_{cons}$ | $PO$ | $\widehat{m}_{cons}$ | $PO$ | $\widehat{m}_{cons}$ | $PO$ | $\widehat{m}_{cons}$ | $PO$ |
| Smoking cessation | | | | | | | | |
| Design by treatment | 0.56 | 1.27 | 0.56 | 1.27 | 0.81 | 4.26 | 0.82 | 4.56 |
| Lu and Ades | 0.56 | 1.27 | 0.58 | 1.38 | 0.82 | 4.56 | 0.83 | 4.88 |
| Jackson | 0.57 | 1.33 | 0.55 | 1.22 | 0.82 | 4.56 | 0.81 | 4.26 |
| Erectile dysfunction | | | | | | | | |
| Design by treatment, Lu and Ades | 0.14 | 0.16 | 0.14 | 0.16 | 0.36 | 0.56 | 0.36 | 0.56 |

***Prior Setups**: a) $\pi_{cons} = 0.5$ and $\boldsymbol{R} = \boldsymbol{I}$, b) $\pi_{cons} = 0.5$ and $\boldsymbol{R} = g(\boldsymbol{Z}'\boldsymbol{Z})^{-1}\sigma^2$, c) $\pi_{cons} \sim Beta(157,44)$ and $\boldsymbol{R} = \boldsymbol{I}$, d) $\pi_{cons} \sim Beta(157,44)$ and $\boldsymbol{R} = g(\boldsymbol{Z}'\boldsymbol{Z})^{-1}\sigma^2$.*

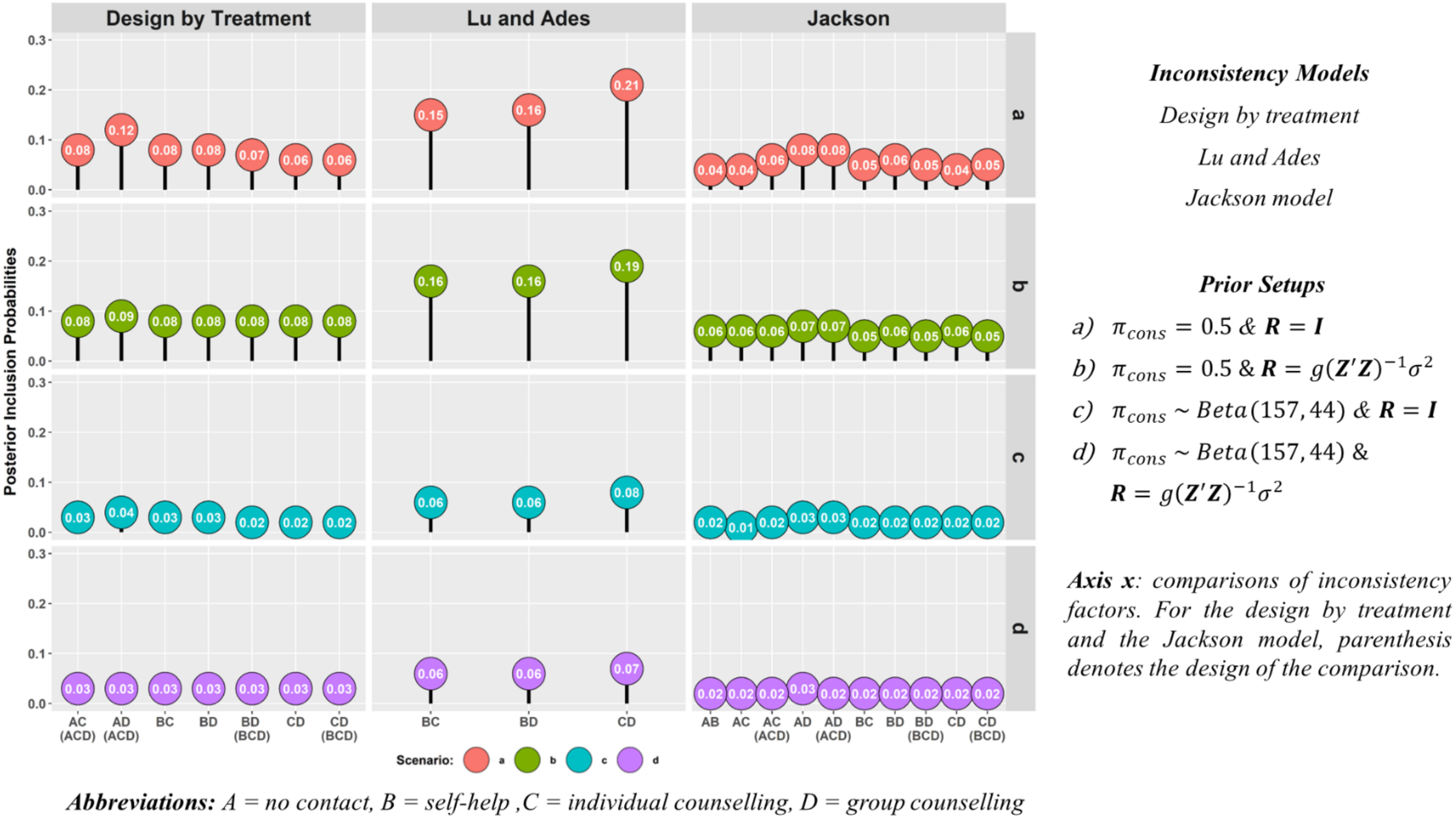

**Figure 4:** Posterior inclusion probabilities for the smoking cessation example for three different inconsistency models.

Note that despite the variability of the posterior inclusion probabilities of each inconsistency factor across different methods, the posterior probability of the consistent NMA model is equivalent and robust, taking values ~0.56 when $\pi_{cons} = 0.5$ and values ~0.82 when the historical based beta prior is used for $\pi_{cons}$. The posterior inclusion probabilities of the Lu and Ades model are larger than the equivalent probabilities of the other models. This is because the Lu and Ades model has fewer inconsistency factors, yielding to smaller prior inclusion probabilities, and the posterior inclusion probabilities in this example are primarily affected by the prior distribution. For example, in the case where $\pi_{cons} = 0.5$, the Lu and Ades model has three inconsistency factors yielding a prior inclusion probability equal to $\pi_{\ell} = 1 - 0.5^{\frac{1}{3}} \approx 1 - 0.79 \approx 0.21$, while the design-by-treatment model has seven with a prior inclusion probability of $\pi_{\ell} = 1 - 0.5^{\frac{1}{7}} \approx 1 - 0.91 \approx 0.09$.

### 3.2. Comparative effectiveness of oral phosphodiesterase type-5 inhibitors for erectile dysfunction

The dataset consists of 69 studies two-arm studies comparing the effectiveness of six type-5 inhibitors (Mirodenafil, Sildenafil, Avanafil, Tadalafil, Udenafil, Vardenafil). Placebo was used as reference treatment for our analysis. Network's geometry is presented Figure 5.

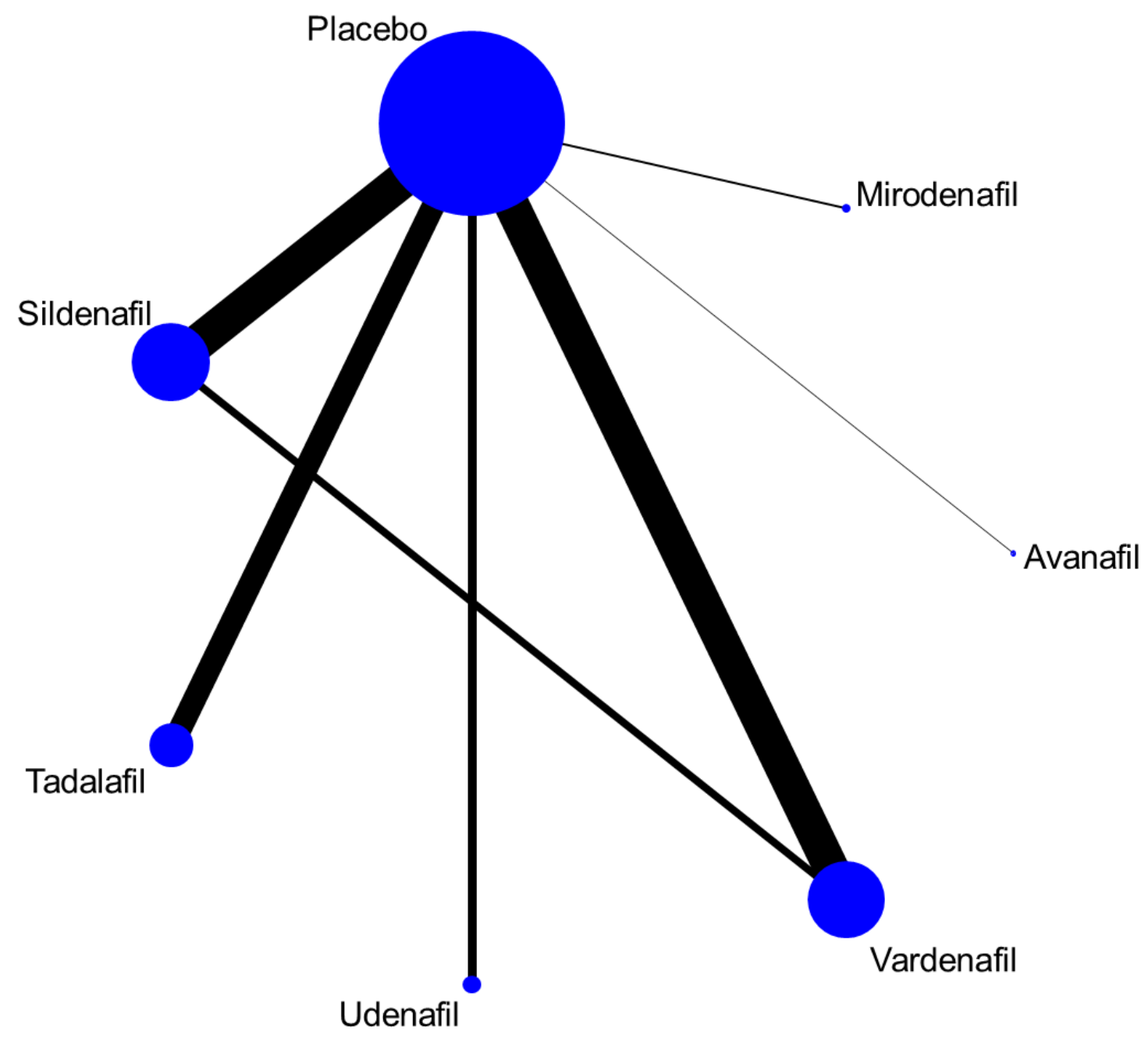

*Edge's thickness and node's size is analogous to the number of treatment comparisons and participants, respectively.*

**Figure 5:** Network geometry of oral phosphodiesterase type 5 inhibitors for erectile dysfunction data.

From the network geometry, we can see that only one closed loop exists between Sildenafil, Vardenafil and Placebo. Because the network contains only two-arms studies, the specification of matrix $\boldsymbol{Z}$ is identical for both design-by-treatment and the Lu and Ades models; only one inconsistency factor can be applied between Sildenafil and Vardenafil. Thus, SSIFS will be the same for these models. For the same reason, the Jackson's model was not used for the specification of matrix $\boldsymbol{Z}$.

Figure 6 indicates that the NMA model is inconsistent since loop inconsistency was identified between treatments Sildenafil, Vardenafil and Placebo. The estimated inconsistency factor's effect is evident when $\pi_{cons} = 0.5$, yielding to a posterior inclusion probability above 0.86 (larger than 0.5, for more details see Table S14) and posterior odds of 6.14 indicating substantial/positive evidence in favor of the hypothesis of inconsistency according to Kass and Raftery (1995)[35]. Since only one inconsistency factor is added in the NMA model, the posterior inclusion probability is equal to the posterior model probability of the inconsistent NMA model. Moreover, the posterior odds show strong evidence of the inconsistent NMA model over the consistent NMA model (for more details see Table 1). Hence, the SSIFS suggests that the network is globally inconsistent due to the loop inconsistency. The inconsistency hypothesis is also supported when the historical-based prior is used for $\pi_{cons}$ (see Tables 1 and S14). Nevertheless, the evidence is much weaker with posterior probability of inconsistency equal to $1 - 0.36 = 0.64$ and posterior odds equal to $1/0.56 \approx 1.78$ indicating marginal evidence in favor of inconsistency. This weaker evidence in comparison with the first "non-informative" approach is logical here since the historical-based prior for $\pi_{cons}$ a-priori supports the consistency hypothesis with prior mean probability equal to 0.78. Having this in mind, we can say that our findings are in general agreement with the classical inconsistency tests and previous work which also suggests that the network is inconsistent. In addition, the Separate Indirect from Direct Design Evidence (SIDDE) method indicates that there is considerable difference between direct and indirect evidence (see Table S15), and the global Q test of inconsistency suggests the presence of global inconsistency ($X_1^2 = 8.44,\ p = 0.004$). Yuan et al. also report that the Bucher test of inconsistency indicates the presence of inconsistency for the comparison sildenafil versus vardenafil (p-value < 0.000)[34].

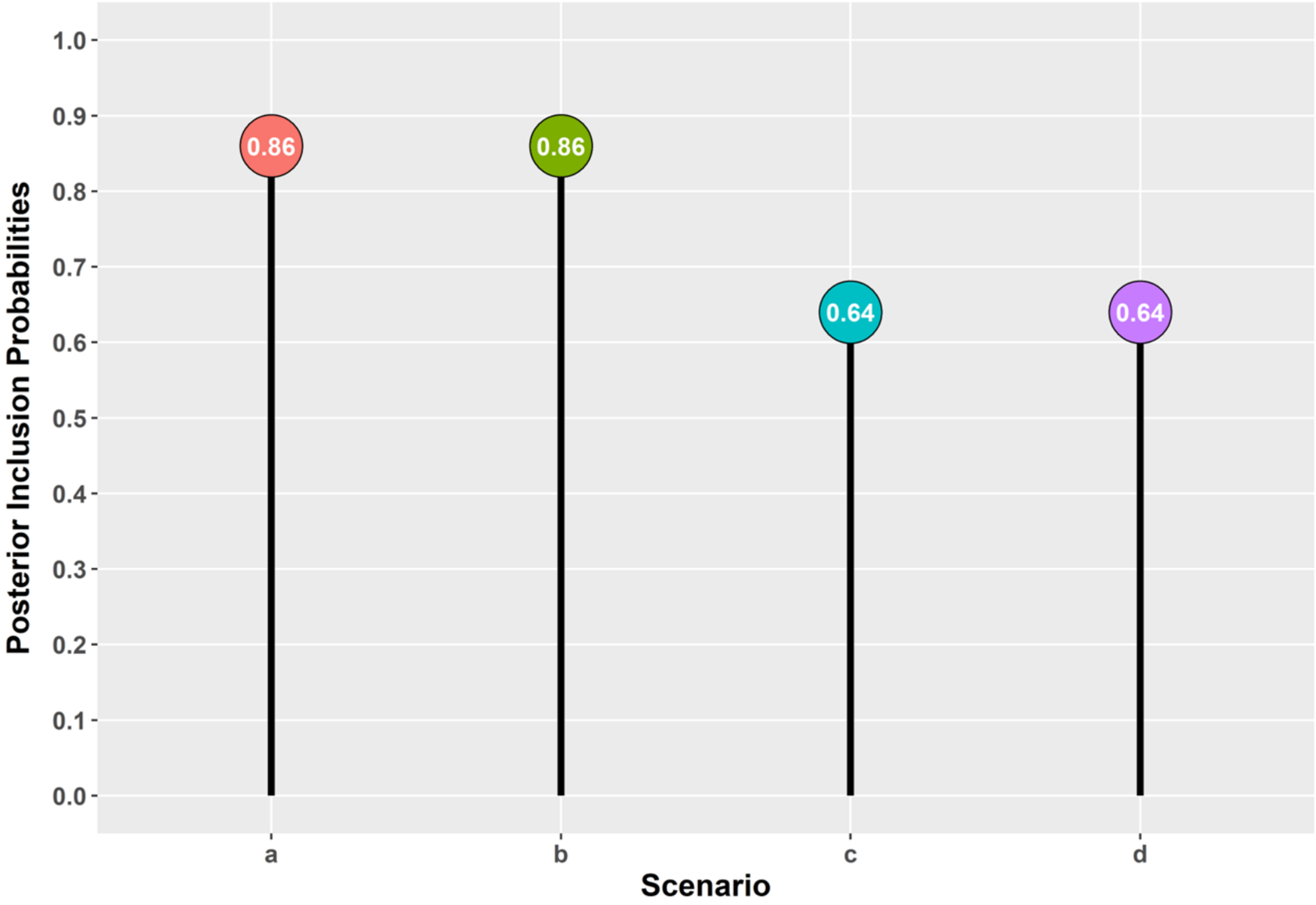

***Horizontal axis***: *4 prior set-ups: a)* $\pi_{cons} = 0.5$ *and* $\boldsymbol{R} = \boldsymbol{I}$, *b)* $\pi_{cons} = 0.5$ *and* $\boldsymbol{R} = g(\boldsymbol{Z}'\boldsymbol{Z})^{-1}\sigma^2$, *c)* $\pi_{cons} \sim Beta(157, 44)$ *and* $\boldsymbol{R} = \boldsymbol{I}$, *d)* $\pi_{cons} \sim Beta(157, 44)$ *and* $\boldsymbol{R} = g(\boldsymbol{Z}'\boldsymbol{Z})^{-1}\sigma^2$.

**Figure 6:** Posterior inclusion probabilities of the inconsistency factor added in the comparison between treatments Sildenafil and Vardenafil for the erectile dysfunction example.

## 4. Discussion

Consistency is a fundamental assumption of NMA, and it should be evaluated correctly to prevent resulting in false conclusions. In this paper, we have viewed the NMA consistency problem as a variable selection problem. Under this perspective, we have implemented Bayesian variable selection via stochastic search variable selection algorithm in NMAs which was named Stochastic Search Inconsistency Factor Selection. SSIFS includes not only the implementation of the Gibbs based SSVS algorithm in NMAs but also the appropriate prior specification of the model parameters and the use of two alternative prior approaches for the inconstancies in our NMA (non-informative and based on historical information).

A common practice for testing consistency in the Bayesian framework, is to compare the consistent NMA model with the fully inconsistent NMA model using the DIC[5,12]. DIC is a function of model deviance and a measure of effective number of parameters ($p_D$). Although, DIC has been criticized for its limitations from several authors[36–39], its usage in NMA for model selection has been widely advocated. Limitations of DIC include concerns about discriminatory performance, invariant $p_D$ to reparameterization, lack of consistency, not based on a proper predictive criterion and weak theoretical justification[39]. Moreover, Millar noticed that DIC based on the conditional likelihood was invalid for hierarchical models[38]. A further limitation of the DIC is the absence of a threshold for judging important differences among models. Moreover, in our problem, DIC can be used only to compare two competing models at each time. Here, we may compare the consistent NMA with the full or a partially inconsistent NMA model. No overall evaluation metric for comparing consistent versus inconsistent models can be obtained in an obvious way. This is in contrast with our proposed approach in this article which (a) estimates the overall posterior weight of the hypothesis of consistency versus inconsistency and (b) important prior information from previous studies can be incorporated in the analysis. By this way, we can adjust our analysis for the multiple comparisons we consider here since many partially inconsistent models are included in our analysis and the evaluation of the consistency hypothesis.

So far, Bayesian tests for inconsistency evaluate the consistency assumption globally. SSIFS evaluates consistency assumption both globally and locally. If for example, we observe a posterior inclusion probability above 0.5 for an inconsistency factor or an inconsistent model with a large posterior model probability, then this is clear evidence in favour of the existence of local inconsistency in our network which indirectly causes global inconsistency in the network. Also, in SSIFS consistency could be evaluated by taking into consideration previous knowledge about how likely is to have a consistent network, by applying an informative prior for $\pi_{cons}$. An informative prior for the probability of having a consistent network is proposed based on a review of 201 networks.

SSIFS also has the advantage that it gives us the option to evaluate the consistency assumption based on a priori difference between direct and indirect evidence that is of practical significance. This difference is depending on the outcome's nature. In continuous outcomes, for example, one unit difference between direct and indirect may be insignificant while in dichotomous outcomes a difference of this magnitude in log-scale is significant. This could be performed in SSIFS, by specifying properly the parameters $c$ and $\psi_{\ell}$.

Extension of this work includes the use of other variable selection methods such as the Gibbs variable selection[40,41] or the Bayesian LASSO[42]. Our proposed method could be implemented through our R package *ssifs* which can be found at "https://github.com/georgiosseitidis/ssifs".

## Appendix

### A) Algorithm for the implementation of the Lu and Ades model

We take the following steps:

- Firstly, exclude the disconnected nodes (if they exist) so that we have a connected network.
- Assume that inconsistency factors could be added in all functional parameters
- Specify all closed-loops of the network by taking all the possible paths that start and end in the same node.
- Check if multi-arm studies are present in the network.
- If multi-arm studies are present: Exclude inconsistency factors from functional parameters present only in one multi-arm trial (that are by definition consistent)
- Exclude the bridges of the network by checking which functional parameters are not included in any closed loop.
- Identify independent loops. Note: A loop is considered independent if it contains at least one edge which is not a part of any other independent loop.
- Keep only the functional parameters that belong to independent loops. Note: Each independent loop must contribute one inconsistency factor, thus for each independent loop we keep only one functional parameter.
- The remaining functional parameters are the comparisons in which inconsistency factor should be added.

**B) Tunning based on assuming a priori the minimum value of inconsistency that is of practical significance**

Tunning parameters $\psi_\ell$ and $c$ in order inconsistency factor to be included in the NMA model when $|b_\ell|$ is larger than a threshold. In the case where the prior correlation matrix is $\boldsymbol{R} = \boldsymbol{I}_p$ we have that

$$f(b_\ell|\gamma_\ell = 0) = f(b_\ell|\gamma_\ell = 1)$$

$$\frac{1}{\psi_\ell\sqrt{2\pi}} e^{-\frac{1}{2}\left(\frac{b_\ell}{\psi_\ell}\right)^2} = \frac{1}{c\psi_\ell\sqrt{2\pi}} e^{-\frac{1}{2}\left(\frac{b_\ell}{c\psi_\ell}\right)^2}$$

$$-\frac{1}{2}\left(\frac{b_\ell}{\psi_\ell}\right)^2 = -log\,(c) - \frac{1}{2}\left(\frac{b_\ell}{c\psi_\ell}\right)^2$$

$$c^2 b_\ell^2 = 2c^2 {\psi_\ell}^2\, log(c) + b_\ell^2$$

$$b_\ell^2(c^2 - 1) = 2c^2 {\psi_\ell}^2 log\,(c)$$

$$b_\ell^2 = \frac{2c^2\psi_\ell^2\, log\,(c)}{c^2 - 1}$$

$$b_\ell^2 = {\psi_\ell}^2 \xi(c), \qquad \xi(c) = \frac{2c^2\, log\,(c)}{c^2 - 1}$$

Inconsistency factors with coefficient $|b_\ell| > \psi_\ell\sqrt{\xi(c)}$ will have larger density to be included in the NMA model. If we assume that the elements of $\boldsymbol{b}$ are correlated then,

$$f(\boldsymbol{b}|\boldsymbol{\gamma} = \boldsymbol{0}) = f(\boldsymbol{b}|\boldsymbol{\gamma} = \boldsymbol{1})$$

$$\left|2\pi\, \boldsymbol{D}_{\boldsymbol{\gamma}=\boldsymbol{0}}\, \boldsymbol{R}\boldsymbol{D}_{\boldsymbol{\gamma}=\boldsymbol{0}}\right|^{-1/2} e^{-\frac{1}{2}\boldsymbol{b}'\left(\boldsymbol{D}_{\boldsymbol{\gamma}=\boldsymbol{0}}\boldsymbol{R}\boldsymbol{D}_{\boldsymbol{\gamma}=\boldsymbol{0}}\right)^{-1}\boldsymbol{b}}$$

$$= \left|2\pi\, \boldsymbol{D}_{\boldsymbol{\gamma}=\boldsymbol{1}}\, \boldsymbol{R}\boldsymbol{D}_{\boldsymbol{\gamma}=\boldsymbol{1}}\right|^{-1/2} e^{-\frac{1}{2}\boldsymbol{b}'\left(\boldsymbol{D}_{\boldsymbol{\gamma}=\boldsymbol{1}}\boldsymbol{R}\boldsymbol{D}_{\boldsymbol{\gamma}=\boldsymbol{1}}\right)^{-1}\boldsymbol{b}}$$

We have

$$\boldsymbol{D}_{\boldsymbol{\gamma}=\boldsymbol{1}} = \begin{pmatrix} c\psi_1 & & \\ & \ddots & \\ & & c\psi_p \end{pmatrix} = c\begin{pmatrix} \psi_1 & & \\ & \ddots & \\ & & \psi_p \end{pmatrix} = c\boldsymbol{D}_{\boldsymbol{\gamma}=\boldsymbol{0}}$$

Thus,

$$\left|\boldsymbol{D}_{\gamma=0}\,\boldsymbol{R}\boldsymbol{D}_{\gamma=0}\right|^{-\frac{1}{2}} e^{-\frac{1}{2}\boldsymbol{b}'\left(\boldsymbol{D}_{\gamma=0}\boldsymbol{R}\boldsymbol{D}_{\gamma=0}\right)^{-1}\boldsymbol{b}}$$

$$= (c^2)^{-\frac{p}{2}}\left|\,\boldsymbol{D}_{\gamma=0}\,\boldsymbol{R}\boldsymbol{D}_{\gamma=0}\right|^{-\frac{1}{2}} e^{-\frac{1}{2c^2}\boldsymbol{b}'\left(\boldsymbol{D}_{\gamma=0}\boldsymbol{R}\boldsymbol{D}_{\gamma=0}\right)^{-1}\boldsymbol{b}}$$

$$e^{-\frac{1}{2}\boldsymbol{b}'\left(\boldsymbol{D}_{\gamma=0}\boldsymbol{R}\boldsymbol{D}_{\gamma=0}\right)^{-1}\boldsymbol{b}} = c^{-p}\, e^{-\frac{1}{2c^2}\boldsymbol{b}'\left(\boldsymbol{D}_{\gamma=0}\boldsymbol{R}\boldsymbol{D}_{\gamma=0}\right)^{-1}\boldsymbol{b}}$$

$$-\frac{1}{2}\,\boldsymbol{b}'\left(\boldsymbol{D}_{\gamma=0}\boldsymbol{R}\boldsymbol{D}_{\gamma=0}\right)^{-1}\boldsymbol{b} = -p\,log\,c - \frac{1}{2c^2}\,\boldsymbol{b}'\left(\boldsymbol{D}_{\gamma=0}\boldsymbol{R}\boldsymbol{D}_{\gamma=0}\right)^{-1}\boldsymbol{b}$$

$$\boldsymbol{b}'\left(\boldsymbol{D}_{\gamma=0}\boldsymbol{R}\boldsymbol{D}_{\gamma=0}\right)^{-1}\boldsymbol{b}\left(\frac{1}{c^2}-1\right) = -2p\,log(c)$$

$$\boldsymbol{b}'\left(\boldsymbol{D}_{\gamma=0}\boldsymbol{R}\boldsymbol{D}_{\gamma=0}\right)^{-1}\boldsymbol{b} = \frac{2pc^2\,log(c)}{c^2-1}$$

Zellner $g$-prior is assumed for the prior correlation matrix $\boldsymbol{R}$

$$\boldsymbol{R} = g(\boldsymbol{Z}'\boldsymbol{Z})^{-1}\sigma^2$$

Hence,

$$\boldsymbol{b}'\left[\boldsymbol{D}_{\gamma=0}(\boldsymbol{Z}'\boldsymbol{Z})^{-1}\boldsymbol{D}_{\gamma=0}\right]^{-1}\,\boldsymbol{b} = \frac{2gp\sigma^2c^2\,log(c)}{c^2-1}$$

Thus, it is considered unsignificant when

$$\boldsymbol{b}'\left[\boldsymbol{D}_{\gamma=0}(\boldsymbol{Z}'\boldsymbol{Z})^{-1}\boldsymbol{D}_{\gamma=0}\right]^{-1}\,\boldsymbol{b} \le gp\sigma^2\xi(c)$$